\documentclass[journal=jacsat,manuscript=article]{achemso}
\usepackage{graphicx}% Include figure files
\usepackage{dcolumn}% Align table columns on decimal point
\usepackage{bm}% bold math
\newcommand{\markup}{}

\author{M. Neek-Amal}
\affiliation [Universiteit Antwerpen] {Departement Fysica, Universiteit Antwerpen, Groenenborgerlaan 171, B-2020 Antwerpen, Belgium}
\alsoaffiliation [Shahid Rajaee Teacher Training University] {Department of Physics, Shahid Rajaee Teacher Training University, Lavizan, Tehran 16785-136, Iran}
\phone{0032 3 265 36 61}
\email{neekamal@srttu.edu, (Tel: +32 3 265 36 61)}
\author{J. Beheshtian}
\affiliation [Universiteit Antwerpen] {Departement Fysica, Universiteit Antwerpen, Groenenborgerlaan 171, B-2020 Antwerpen, Belgium}
\alsoaffiliation [Shahid Rajaee Teacher Training University] {Department of Physics, Shahid Rajaee Teacher Training University, Lavizan, Tehran 16785-136, Iran}
\phone{0032 3 265 36 61}
\author{A. Sadeghi}
\author{K. H. Michel}
\affiliation [Universiteit Antwerpen] {Departement Fysica, Universiteit Antwerpen, Groenenborgerlaan 171, B-2020 Antwerpen, Belgium}
\author{F. M. Peeters}
\affiliation [Universiteit Antwerpen] {Departement Fysica, Universiteit Antwerpen, Groenenborgerlaan 171, B-2020 Antwerpen, Belgium}

\date{\today}
\title{Boron Nitride Monolayer: A Strain Tunable Nanosensor}

\begin{document}

\break

\begin{abstract}

The influence of triaxial in-plane strain on the electronic
properties of a hexagonal boron-nitride sheet is investigated using
density functional theory. Different from graphene, the triaxial
strain localizes the molecular orbitals of the boron-nitride flake
in its center depending on the direction of the applied strain. The
proposed technique for localizing the molecular orbitals that are
close to the Fermi level in the center of boron nitride flakes can
be used to actualize engineered nanosensors, for instance, to
selectively detect gas molecules.  We show that the central part of
the strained flake adsorbs polar molecules more strongly as compared
to an unstrained sheet.

\end{abstract}

%\keywords

%%%%%%%%%%%%%%%%%%%%%%%%%%%%%%%%%%%%%%%%%%%%%%%%%%%%%%%%%

\maketitle

\break

\section{Introduction}
Strain engineering can be used to control the electronic properties
of nanomaterials. This is of interest for fundamental physics, but
is also relevant for potential device applications in
nanoelectronics. Because the electronic and mechanical properties of
an atomic monolayer are strongly influenced by strain they have
attracted considerable attention over the last
decades~\cite{3,naturephys2010}.

Unlike graphene, a h-BN sheet is a wide gap insulator, as is bulk
h-BN, and  is a promising material for opto-electronic technologies
~\cite{14,15,zettle2008}, tunnel devices and field-effect
transistors~\cite{naonoletter2012}. Using a combination of
mechanical exfoliation and reactive ion etching, monolayer and
multilayer suspended h-BN sheets can be prepared~\cite{PRB2009US}.
The band gap of boron nitride nanoribbons can be altered by edge
passivation with different types of atoms~\cite{23,25,26}.

A combination of an odd number of h-BN  layers is a non-centrosymmetric
ionic crystal which is piezoelectric due to
 $D_{3h}$ symmetry~\cite{Karl2009,Sai}. The corrugations on the h-BN sheet results in a strong polarization
in the plane of the sheet which depends non-analytically on the wave
vector of the corrugations~\cite{PRL2009}. \markup{h-BN sheet has a
non-linear elastic deformation up to an ultimate strength followed
by a strain softening to failure~\cite{peng1,peng2}.} Moreover, the
band gap of boron nitride nanotubes can be reduced by a transverse
electric field due to a mixing of states within the highest occupied
molecular orbital and the lowest unoccupied molecular
orbital~\cite{PRBJavad2012,PRB69,21,Zettle2005}. \markup{ The
reduction in the band gap due to uniaxial strain results in
tunneling magnetoresistance ratio which increases linearly with
applied strain~\cite{jchem}.} Here we propose an alternative
approach for electron hole localization based on a tunable
parameter, i.e. inhomogeneous strain.

Developments of nanosensors of (different) gases is to a great
extent related to the response to both the morphology and the
surface states of the material. Single-wall carbon nanotubes (SWNT)
can act as a chemical sensor for sensing gaseous molecules such as
NO$_2$ or NH$_3$ where the electrical resistance of a semiconducting
SWNT is found to dramatically increase or
decrease~\cite{Zettle1997,Zettle2000,science2000,Review}. Here we
study the effect of strain on the adsorption mechanism and propose a
new and tunable way to control the adsorption of a gas.

In this paper, using density functional theory (DFT)
calculations, we show a spatial separation of the
highest occupied and lowest unoccupied molecular orbitals (i.e. HOMO and LUMO)
in response to a triaxial in-plane strain.
The result is in agreement with the predictions from
piezoelectricity theory. Consequently, the binding energy of an
external polar molecule over the strained sample is considerably
enhanced. Depending on the applied triaxial strain on the zig-zag
edges with boron (nitrogen) termination the HOMO (LUMO) is confined
in the central portion of the flake. This study opens a new avenue
in the field of strain engineering of a monolayer of h-BN in terms
of tunable spatial localization of the frontier orbitals. (it
controls and enhances  chemical reaction). In
 recent experiments the edge structure of graphitic
nanostructures were successfully controlled~\cite{science323} and
well defined (e.g. hexagonal shape) graphene flakes with zig-zag
edges were observed~\cite{PRL2011}. Consequently, the proposed
 experimental set up is realistic and therefore we expect that the calculated effects will be measurable on
micron size samples employing experimental realized controlled edge
chirality~\cite{nanolett2011,science323,PRL2011}. A simple
experimental set up for creating triaxial strain was proposed in
Ref.~\cite{naturephys2010}.

The paper is organized as follows. In Sec. II, we present our
theoretical approach for triaxial strain and  corresponding
piezoelectricity. In Sec. III we present the molecular dynamics
simulation and density functional calculation methods. Then in Sec.
IV we give and discuss our results. We conclude the paper in Sec. V.

\begin{figure}
\begin{center}
\includegraphics[width=0.425\linewidth]{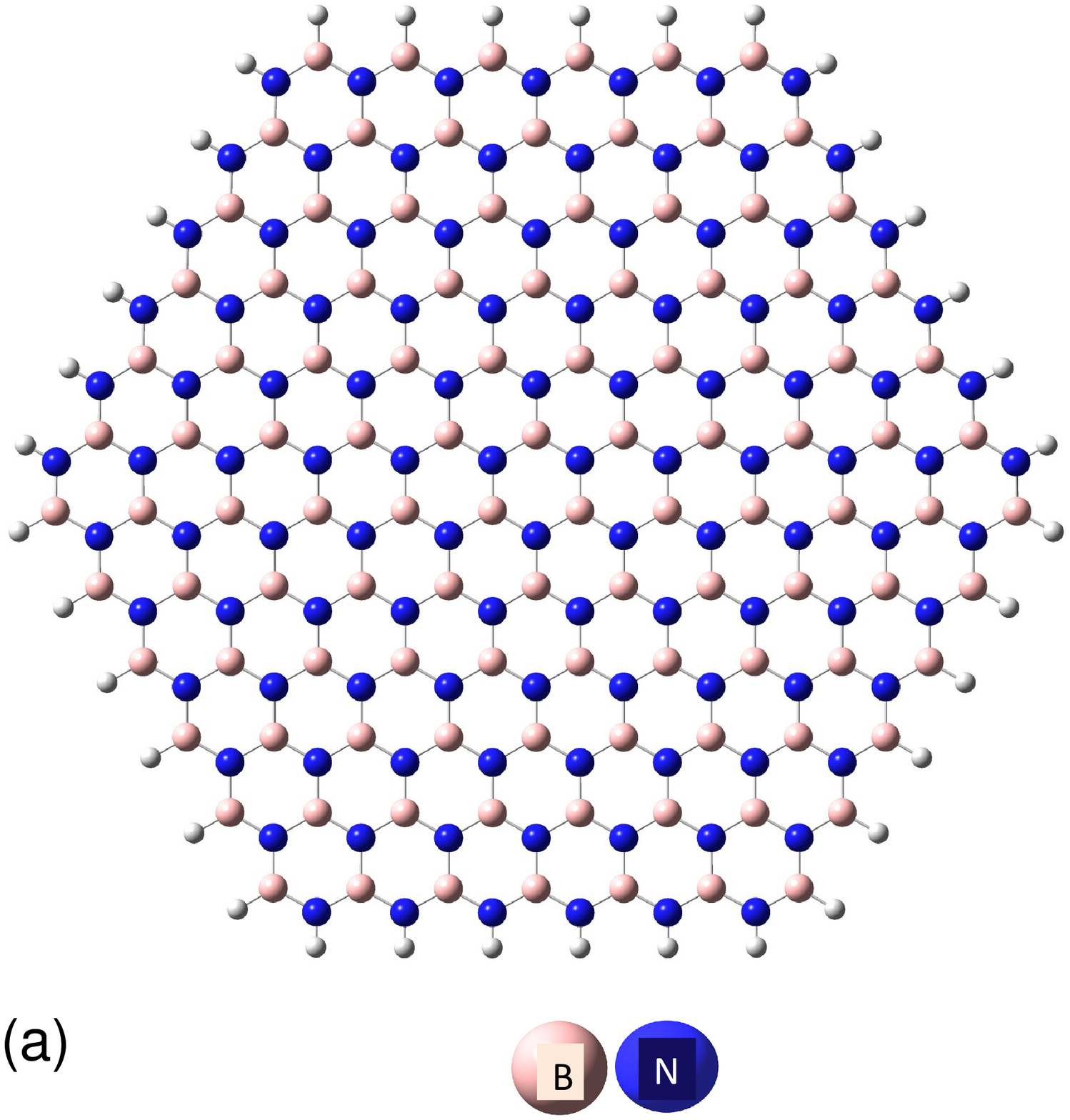}
\includegraphics[width=0.425\linewidth]{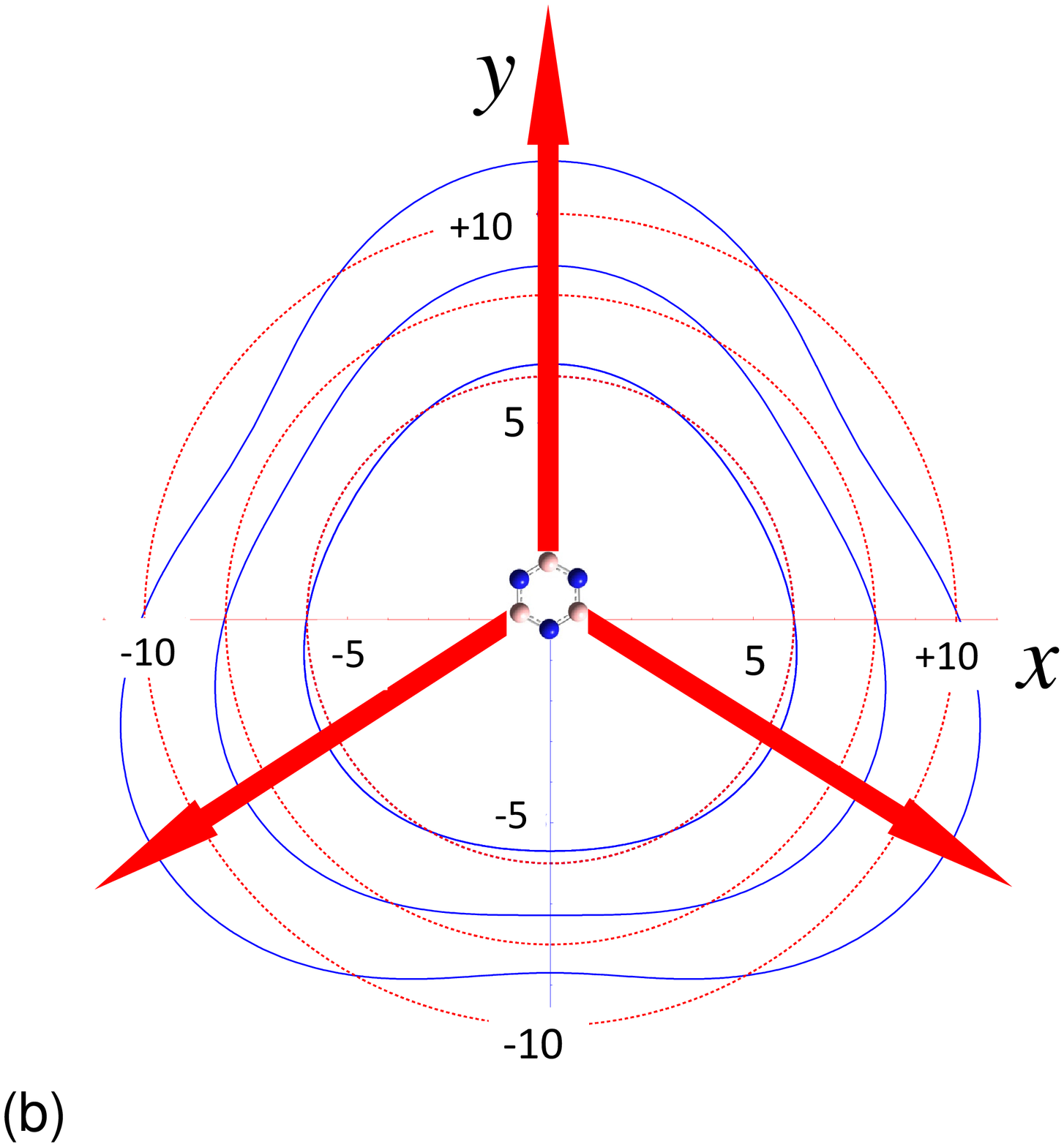}
\caption{  (a) Hexagonal flake of BN passivated by  H
atoms (white balls). (b) A schematic representation of the distorted
hexagonal boron nitride flake by the applied triaxial strain. The
red curves represent the original shape and the blue curves indicate
the distorted flake. The flake is stretched along the three
crystallographic directions which are represented by the three red
vectors. The NB system is obtained by interchanging B and N atoms,
or equivalently by rotating (a) and the central hexagon in (b)  by
an angle $\pi/3$, see \ref{figstress}(d,e,f). \label{figsketch}}
\end{center}
\end{figure}

\begin{figure}
\begin{center}
\includegraphics[width=0.8\linewidth]{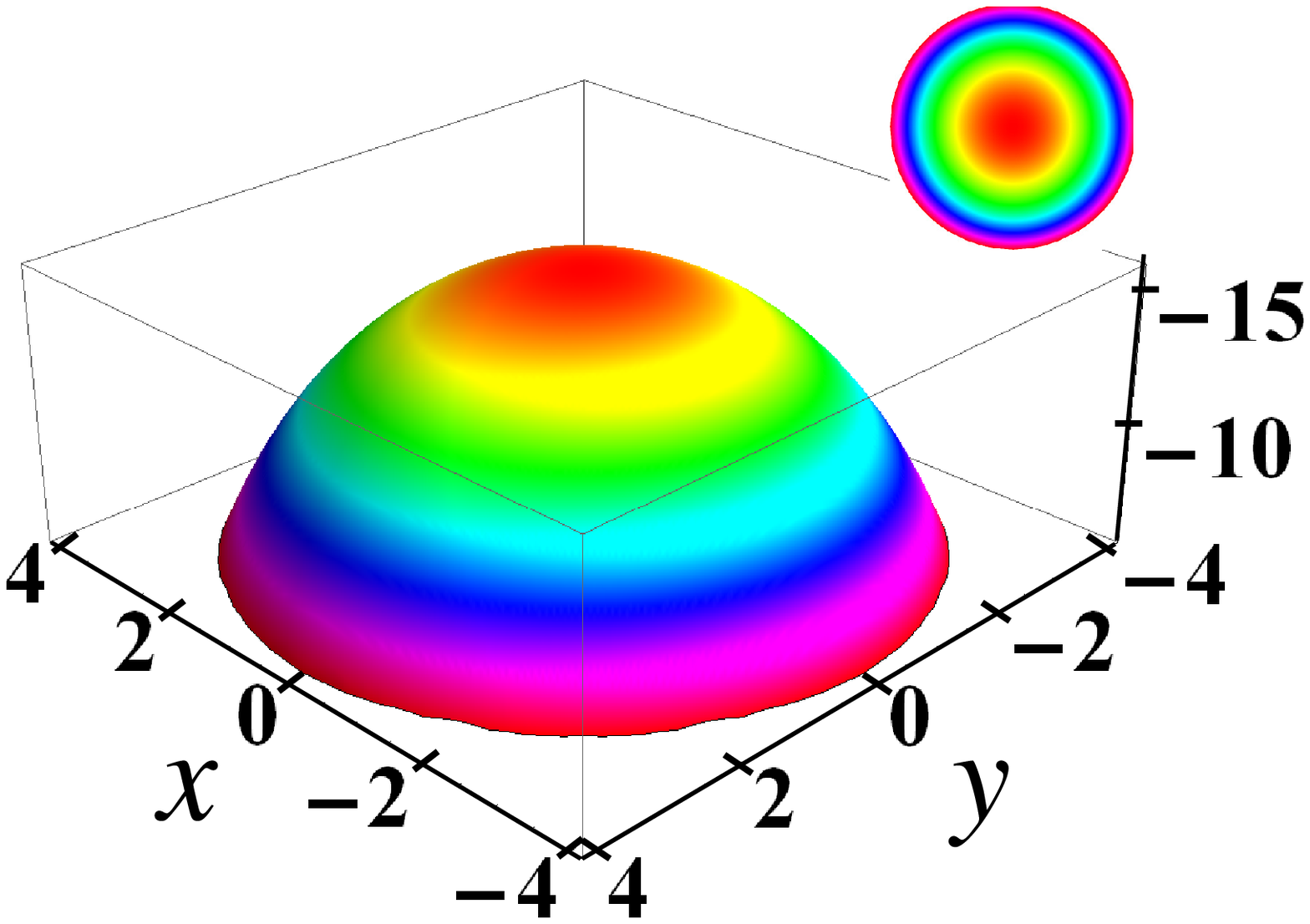}
\caption{  Electrostatic potential in arbitrary  units
% of $2d_0\mu C/\pi\epsilon_0$
at distance $z=2\times10^{-4}R$ above a disk of radius $R$ subject
to triaxial  strain as predicted by our theoretical model based on
 linear theory of piezoelectricity (inset shows top view).
\label{figPot}}
\end{center}
\end{figure}

\section{Theoretical Model}

\ref{figsketch}(a) shows a hexagonal h-BN flake  with zig-zag
edges  passivated by hydrogen. The distortion of a hexagonal
boron-nitride flake subjected to triaxial strain along three
equivalent crystallographic directions is shown schematically in
\ref{figsketch}(b). The original shape is shown by the
dotted-red circles and the deformed shape by the blue-solid curves.
In polar coordinates $(r,\theta)$ the applied triaxial strain
results in a displacement vector~\cite{naturephys2010}
$\markup{u}=(u_r,u_{\theta})=C\,r^2(\sin(3\theta),\cos(3\theta))$,
where $C$ is a constant determining the strength of the applied
strain and has dimension of inverse length. Notice that the
$r^2$-dependence in $\markup{u}$ insures that the applied triaxial
strain can also be realized on an infinite or macroscopic sheet. In
the following we will first present a simple analytical theory that
agrees qualitatively with our numerical DFT results.

Linear elasticity theory for an isotropic material leads to the
stress-strain relation, i.e. $\sigma_{jk}=\lambda
\delta_{jk}\nabla.\markup{u}+2\mu \epsilon_{jk}$, where $\lambda$
and $\mu$ are the Lam\'{e} parameters that determine the stiffness
of the material. If we substitute $\markup{u}$  in the latter
equation the components of the stress tensor in polar coordinates
are written as

~~~~~~~~~~~ $\sigma(r,\theta)= 4\mu\,C\,r\,\left(
  \begin{array}{cc}
   \sin(3\theta) & \cos(3\theta) \\
     \cos(3\theta) &  -\sin(3\theta) \\
  \end{array}
\right).$

Here, it is more convenient to use the components of the stress
tensor in Cartesian coordinates where the $y$ axis is taken along
the arm-chair direction and the $x$ axis is taken along the zig-zag
direction.  The edges under strain can have B (called BN system) or
N (called NB system) atoms (e.g. in \ref{figsketch}(a), the
strain is applied on B atom edges, i.e. the BN system is stretched
along the red arrows). Note that the three strained edges (or free
edges)
%should
have only one type of atoms. We consider here the case of
strained  B edges, i.e. the BN flake of \ref{figsketch}.

Using the product $\sigma(x,y)=\Re \sigma(r,\theta) \Re^T$, where
$\Re$ is the rotation matrix about the $z$ axis, the stress tensor
in Cartesian coordinates can be rewritten as
\[\sigma(x,y)= 4\mu\,C\,\left(
  \begin{array}{cc}
   y & x \\
    x &  -y \\
  \end{array}
\right).\]

On the other hand, an elastic in-plane deformation of the h-BN flake
lowers its lattice symmetry, redistributes the valance charges in
terms of shifting $\sigma$ and $\pi$ bonds and  produces a non-zero
polarization. Using linear piezoelectricity theory, the induced
polarization due to the applied strain can be written as
${P_i}=d_{i,jk}\sigma_{jk}$, where $d$ is the third rank
piezoelectricity tensor which has in general (in two dimensions) 8
elements where the indices $(i,j,k)$ can be $x$ and $y$. The 3\,$m$
symmetry  of the h-BN sheet results in only one independent element
for the piezoelectricity tensor, $d_0=d_{y,yy}$. The tensor is
invariant under a rotation angle of $2\pi/3$ about the $z$-axis
which yields the following symmetry relations:
$d_{y,yy}=-d_{y,xx}=-d_{x,yx}=-d_{x,xy}$. Substituting $\sigma(x,y)$
in Eq.~(1) results in the local induced dipoles:
\begin{equation}
P_x=-8\,d_0\,\mu\,C\,x,~~~~~~~~ P_y=-8\,d_0\,\mu\,C\,y .
\end{equation}

\begin{figure*}
\begin{center}
\includegraphics[width=0.7\linewidth]{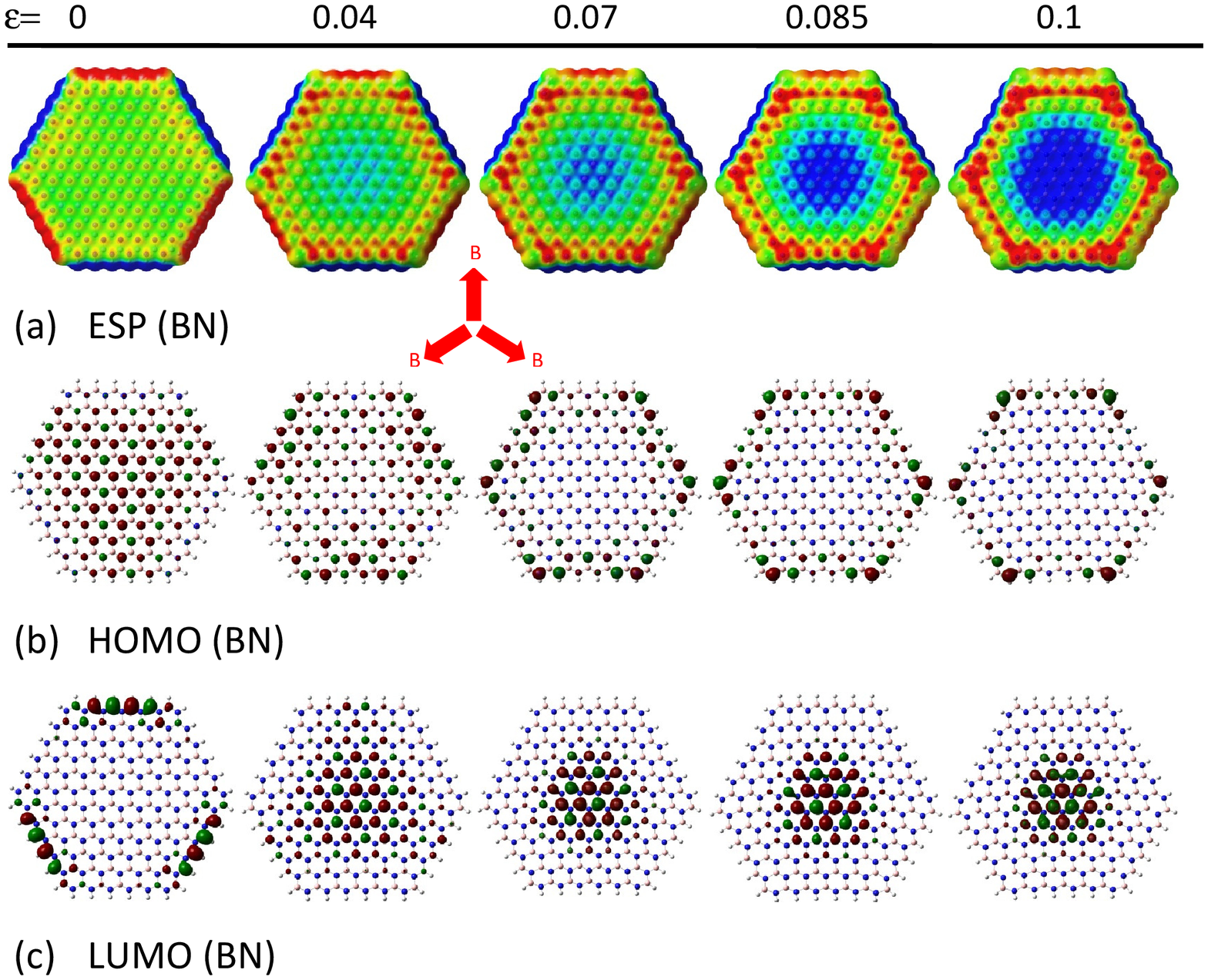}
\includegraphics[width=0.7\linewidth]{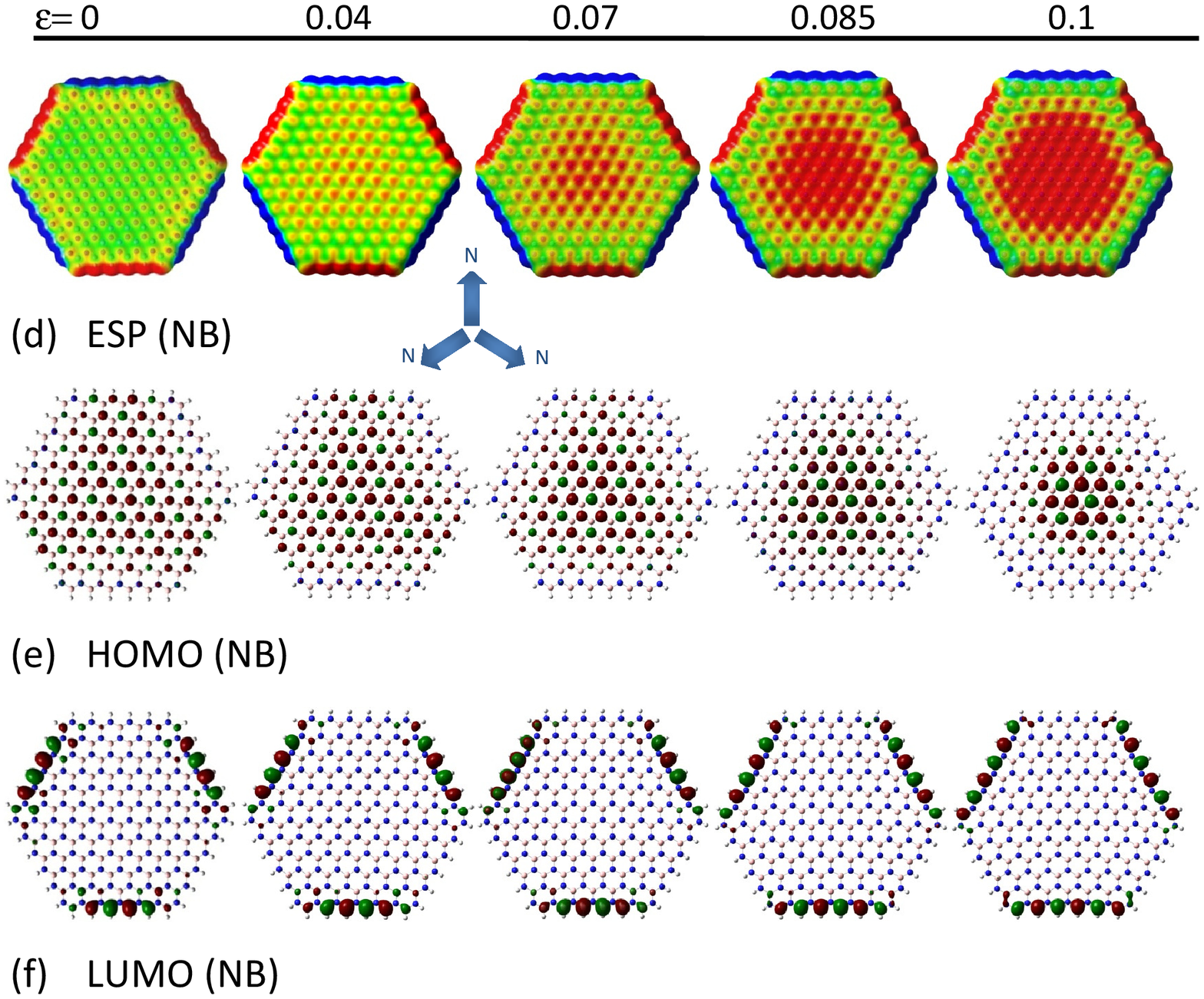}
\caption{  (a,d) Electrostatic  potential surface
mapped on an isosurface of electronic density of
$1.5\times10^{-2}$~e/\AA$^3$,~ (b,e) highest occupied and (c,f)
lowest unoccupied molecular orbitals for hexagonal shaped  h-BN
flakes subject to different triaxial strains $\epsilon$ applied
along the directions indicated by the arrows. The lower (higher) ESP
is indicated by red (blue) color.
 In BN system (a, b, c) the stretched edges are terminated by B atoms while
 in NB system (d, e, f) the edges terminated by N atoms are stretched.
Note that rotating the $\epsilon=0$ panels in (a,b,c) by an angle
$\pi/3$ gives the $\epsilon=0$ panels in (d,e,f), respectively.
 \label{figESP}}
\end{center}
\end{figure*}

Note that the local dipoles are directed radially inward, i.e.
$\markup{P}=-8\,d_0\,\mu\,C\,r {\vec e}_r$ with magnitude
$8\,d_0\,\mu\,C$ per unit of radial distance. For a disk with
diameter $D=2R$ and using $C=\delta/D$, the total induced dipole
moment is found to be zero by integrating over the disk from 0 to
$\phi$
\begin{equation}
{P^{T}_x}=-2\,d_0\,\mu\,\delta\,R\,\sin(\phi),{P^{T}_y}=+2\,d_0\,\mu\,\delta\,R\,\cos(\phi),~~~~~~~~
\end{equation}
where $\phi=2\pi$
 i.e. $P^{T}_x(2\pi)=P^{T}_y(2\pi)$=0. Notice that $P^{T}_i(\pi)=P^{T}_i(-\pi)$. The local $\markup{P}$
results in a surface charge density ($\sigma_p=-\nabla.\markup{P}$)
and a boundary charge density ($\lambda_p=-\markup{P}. {\vec e}_r$),
hence the corresponding electrostatic potential
$\phi_P(\overrightarrow{x})$ which is proportional to $
 \int\frac{\sigma_p
ds'}{|\overrightarrow{x}-\overrightarrow{x}'|}+\oint\frac{\lambda_p
dl'}{|\overrightarrow{x}-\overrightarrow{x}'|} $ can be written in
terms of Bessel functions of the second kind and results in a
radially decreasing potential. For a disk with radius $R$, the first
integral is taken over the disk surface and the second is taken over
its perimeter. In \ref{figPot} we show
$\phi_P(\overrightarrow{x})$
% (in theunit of $2d_0\mu C/\pi\epsilon_0$)
in x-y plane at height $z=2\times10^{-4}R$ above a circular flake
with radius $R$. These results are in qualitative agreement with the
electrostatic potential (ESP) obtained from our DFT results shown in
~\ref{figESP}(a,d). Notice that for uniaxial strain, e.g.
$\markup{u}=(x,0)$ and shear strain, e.g. $\markup{u}=(y,x)$ the
used formalism gives $\markup{P}=(P_x,0)=(\mu\,d_0,0)$ and
$\markup{P}=(0,P_y)=(0,-2\mu\,d_0)$, respectively, which are in
agreement with the DFT results of Ref.~\cite{Sai}. It is important
to note that the above model is size-independent and is valid also
for an infinitely large h-BN flake. On the other hand applying
strain on the N-edges (NB-system) is equivalent to the
transformation $\theta\rightarrow \theta+\pi/3$ in $\markup{u}$
which yields $\markup{u}_{NB}=-C\,r^2(\sin(3\theta),\cos(3\theta))$.
Rewriting the above theory for the latter displacement vector
results in $\markup{P}=8\,d_0\,\mu\,C\,r \overrightarrow{\bf{e_r}}$
which has the opposite direction of the dipole moment of BN. This is
in agreement with our DFT results shown in
~\ref{figESP}(d,e,f) for NB (we will discuss our DFT results
below). We conclude that for an infinite hexagonal flake with
zig-zag edges we have the opposite localization scheme for HOMO and
LUMO depending on whether strain is applied on the N-edges or the
B-edges.

\begin{figure*}
\begin{center}
\includegraphics[width=0.9\linewidth]{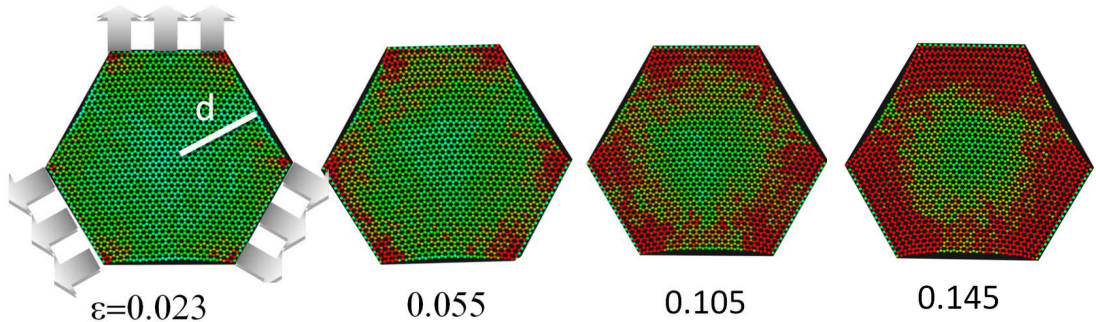}
\caption{  Four representative snapshots of molecular
dynamics simulations of the hexagonal shaped h-BN flake subject to
triaxial strain. The red and blue colors indicate the regions with
highest and lowest stress, respectively. The direction of the
applied strain $\epsilon$ is shown by the arrows in the first
snapshot and $d$=4.6~nm.
 \label{figstress}}
\end{center}
\end{figure*}

\section{Computational Details}
%Moreover,
First, in order to study the stress distribution~\cite{PRBlucian}
on a large scale h-BN flake, we performed classical molecular dynamics (MD)
simulation  at $T$=300~K for a system with 2400 atoms.
%($d_0=3.5$\,nm). Ali says: I put it in caption of the figure
We used a modified Tersoff potential (which is defined  in the
LAMMPS package~\cite{lammps,Plimpton}) using the parameters proposed
by Sevik~\emph{et al}~\cite{Sevik,sandeep2013} for a h-BN sheet. % All the
%parameters and a detailed description of the potential energy are
%given in Ref.~\cite{Sevik}.
%The method for determining the stress on
%each atom can be found in our previous work~\cite{PRBlucian}.
\ref{figstress} shows four snapshots of our MD simulation for
a h-BN flake with three of the zig-zag edges subject to triaxial
strain as
shown by the arrows. % The highest stress is shown by red color and the lowest stress is shown by blue color.
 Notice that the corners
are always under higher stress (red color) while the central atoms
are subject to a reduced stress (blue color) (this results in a high
pseudomagnetic field at the corners in case of
graphene~\cite{naturephys2010}).

To study the electronic behavior of the flakes in response to
triaxial strain, we employ DFT as implemented in the GAUSSIAN (G09)
package~\cite{G09}. The electronic wavefunction is expanded using
the 6-31G* Gaussian type basis set and the exchange-correlation  is
treated using the hybrid functional B3LYP. The self consistency loop
iterates until the change in the total energy is less than
$10^{-7}$~eV. In an unstrained hexagonal flake, distance of the
edges to the center is denoted by $d_0$. The triaxial strain
($\epsilon=\delta/d_0$)
 is applied  stepwise
by rigidly displacing the atoms on the edges %by a small amount
in directions normal to the corresponding edges such that the distance of the edges
 from the center becomes $d=d_0+\delta$.
%Then we keep all atoms on three non-adjacent edges  frozen in their
%new positions while other nuclei (including hydrogens which saturate
%the dangling bonds at the edges) are well relaxed until the force on
%each is less than 50~meV/\AA.
 Note that all atoms on each zig-zag
edge are of the same type (see \ref{figsketch}).

\section{Results and Discussion}

\markup{In order to explain the change in the  main electronic
properties as resulting form triaxial strain in BN and NB flakes one
should consider the following issues: i) contrary to the fixed
edges, the free edges are relaxed (hence  can expand) when applying
strain;
 ii) as we showed in section II both
BN and NB under stress become polarized locally with different
orientation while the total dipole moment is zero, iii) the
polarization and the electrostatic potential distribution in the
system correlate, and iv) all edges (both BN and NB) are passivated
by hydrogens and because of the difference of electronegativity of
H,  B and N the results will be different for unsaturated flake
edges.}

\subsection{Localized states and electrostatic potential}

\begin{figure*}
\begin{center}
\includegraphics[width=0.31\linewidth]{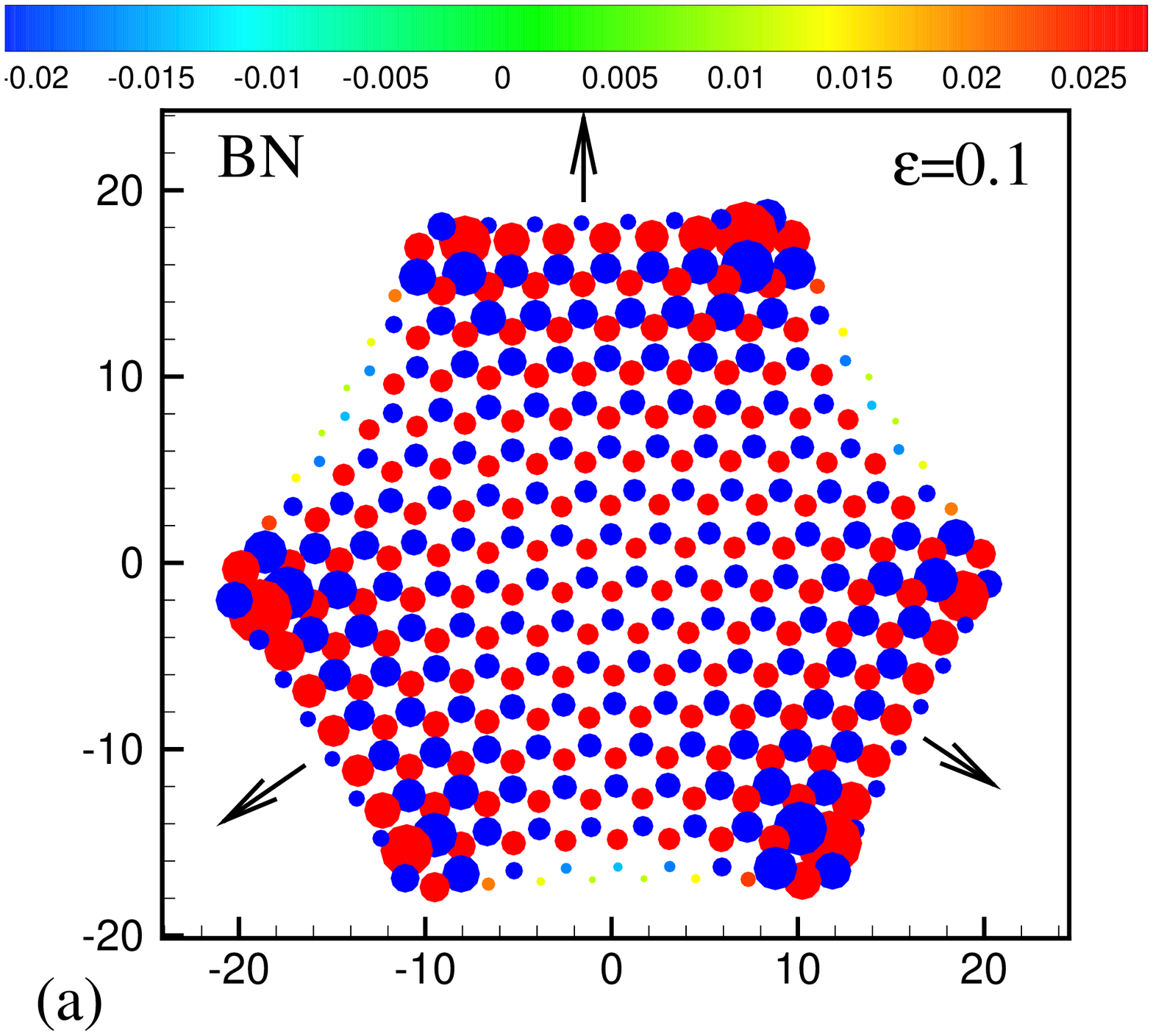}
\includegraphics[width=0.31\linewidth]{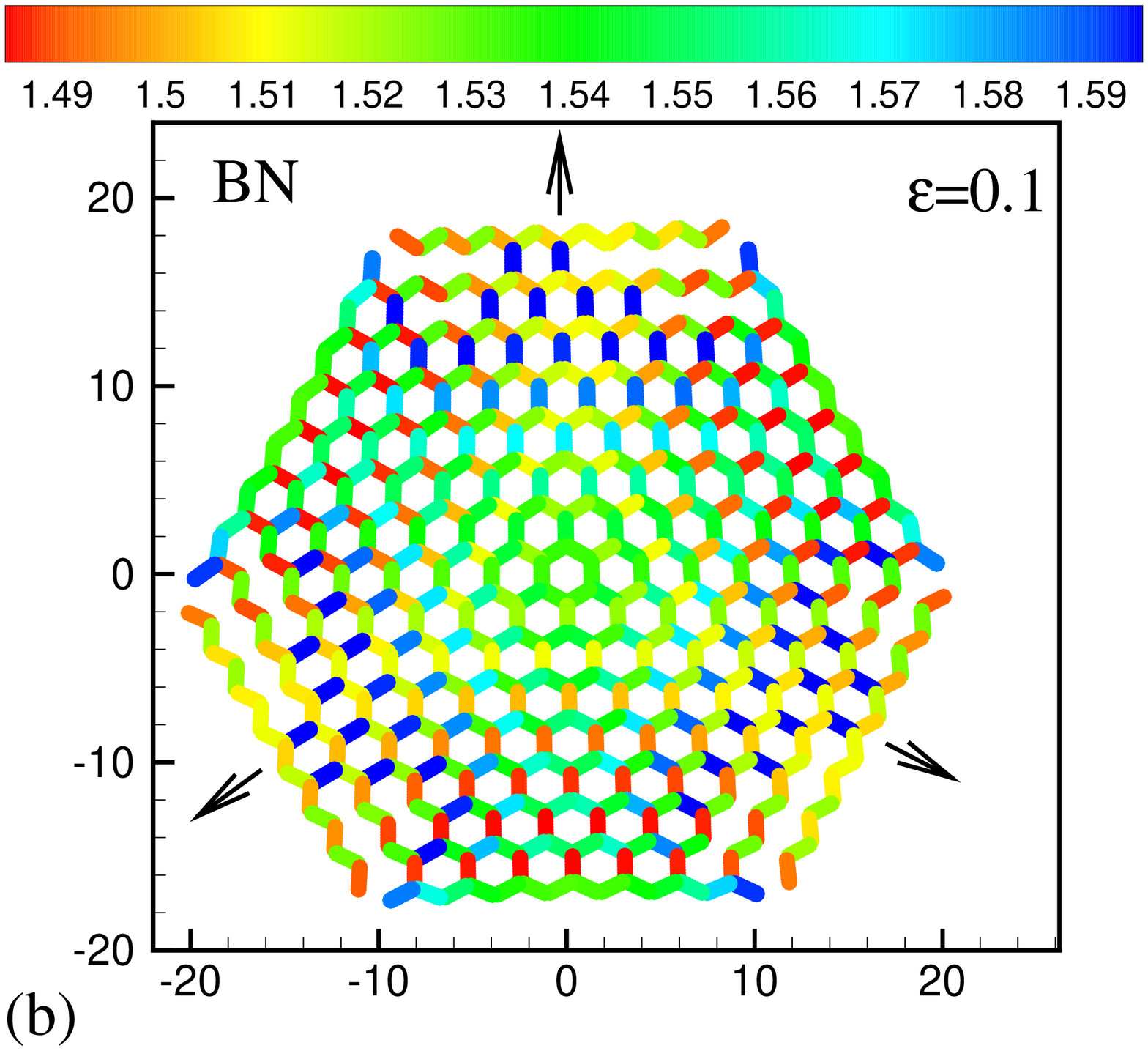}
\includegraphics[width=0.31\linewidth]{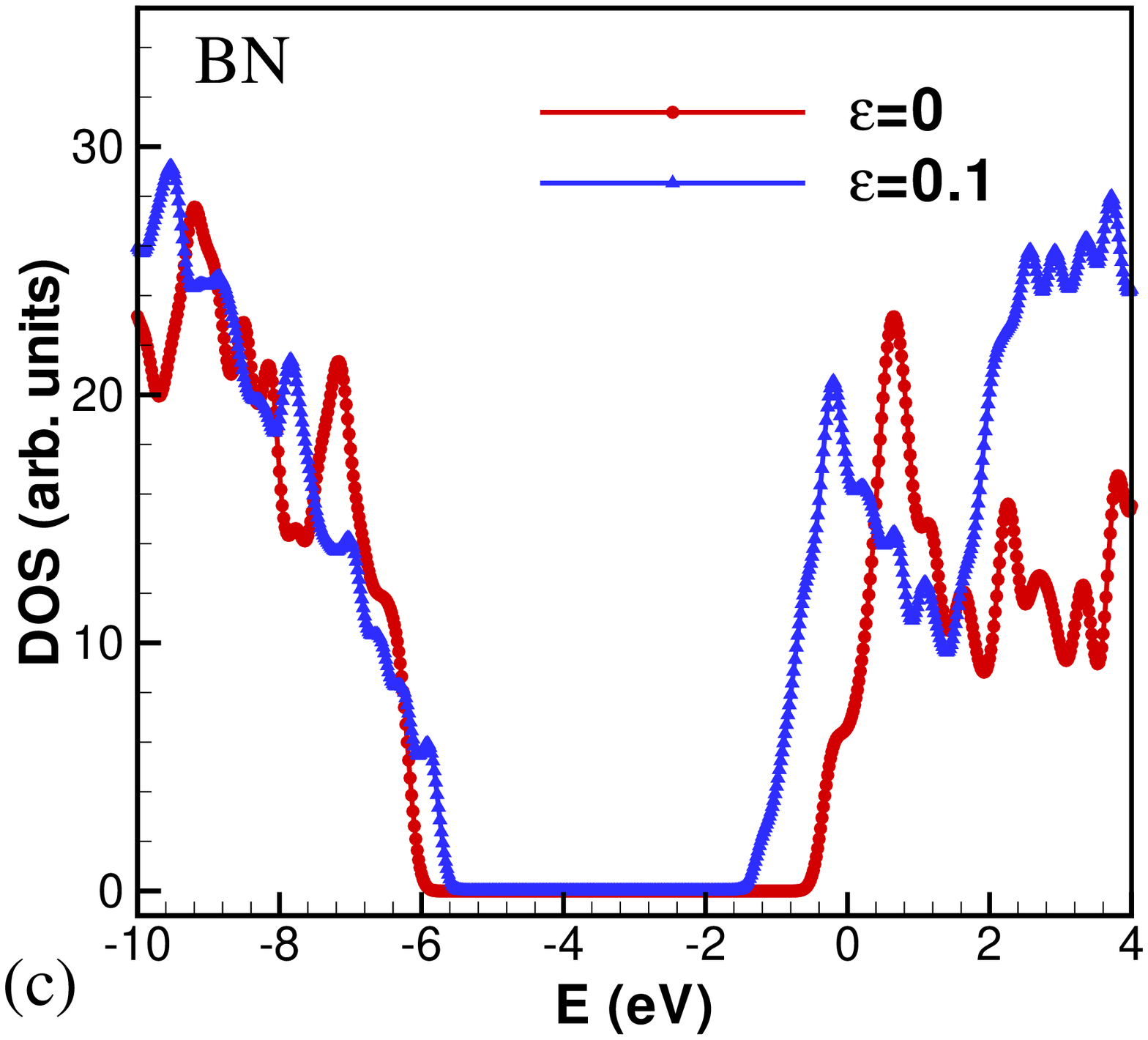}
 \caption{  \label{figNEW}
\markup{(a) The Change in the Mulliken charges induced by applying 
strain  ($\epsilon=10\%$) to the BN flake. Circle radius corresponds to the 
charge difference between strained and unstrained BN flakes.
 (b) The bond length, i.e. $a_{BN}$, distribution in a strained
BN flake ($\epsilon=10\%$). For clarity, bonds longer
than 1.6~\AA~ are not shown.
 (c) DOS spectra of strained (10\%) and unstrained BN flakes.
} }
\end{center}
\end{figure*}

 The electrostatic potential, highest occupied and lowest unoccupied molecular orbitals
obtained from the DFT calculations for a hexagonal shaped  h-BN
flake consisting of 252 atoms
 %(i.e. B$_{108}$N$_{108}$H$_{36}$ with $d_0=1.35$~nm)
 ($d_0=1.35$~nm) 
are shown in \ref{figESP} for five different
values of $\epsilon=\delta/d_0$.
 When the flake is stretched at the B edges (i.e. BN system, \ref{figESP}(a-c)), both the
region with higher ESP and the LUMO  are localized in the central
part. On the other hand,
 when strain is applied on the N edges (i.e. NB system, ~\ref{figESP}(d-f)), the
region with lower ESP and the HOMO both are localized again in the
center. \markup{In the unstrained ($\epsilon$=0) flakes the LUMO is
localized on the N-edges, while the HOMO is not concentrated on the
B-edges which is different from the case of  rectangular
ribbons~\cite{PRBJavad2012,PRB72}. For a rectangular ribbon the N
atoms absorb electrons from H and therefore the HOMO is localized on
the N atoms while in the B edges the H atoms gain electronic charge
from those B atoms and the B atoms loose their electrons and
consequently the LUMO is on the B atoms. In the hexagonal unstrained
flakes, the N atoms absorb electrons and the HOMO is also on the N
atoms. But the LUMO is only on the B atoms in the mid point of the B
edge while the B atoms situated at the corners of the B edge do not
contribute in the LUMO. 
The reason is that the corner B atoms recover their lost electrons from their N neighbors at the same corner. 
In ~\ref{figNEW}(a) the Mulliken charge change when
applying 10$\%$ strain is shown for a BN flake with 432 atoms. It is
seen that the B atoms transfer electrons to N atoms which make the
system highly polarized. The charge distribution reveal the
corresponding triaxial applied stress shown in \ref{figsketch}(b).}

 The larger $\epsilon$, the more localized the HOMO (LUMO) in the
center of the NB (BN) flake. \markup{This is due to the appearance
of longer bond-lengths ($a_{BN}$). The longer the B-N bond length,
there is the less 2$p_z$ hybridization. In \ref{figNEW}(b)
the distribution of $a_{BN}$ for a strained BN flake
($\epsilon=10\%$) is shown 
%(for more clarity we do not show those bonds longer than 1.6~\AA).~
(bonds longer than 1.6~\AA~ are not shown for clarity).
This bond length distribution shows the
weakening of the covalent bond perpendicular to the stressed edges.
One can connect this pattern to those shown in
\ref{figESP}(a,b,c). The longer the bonds perpendicular to
the B (N) edge in a BN (NB) flake the larger the inward (outward)
dipole moment.}

 The gradient in ESP from  the
edges into the center increases with increasing $\epsilon$. We
performed also DFT calculations to study similar effects in  a
graphene flake with the same size but no such
localization/polarization effects were found.

\begin{figure}
\begin{center}
\includegraphics[width=0.7\linewidth]{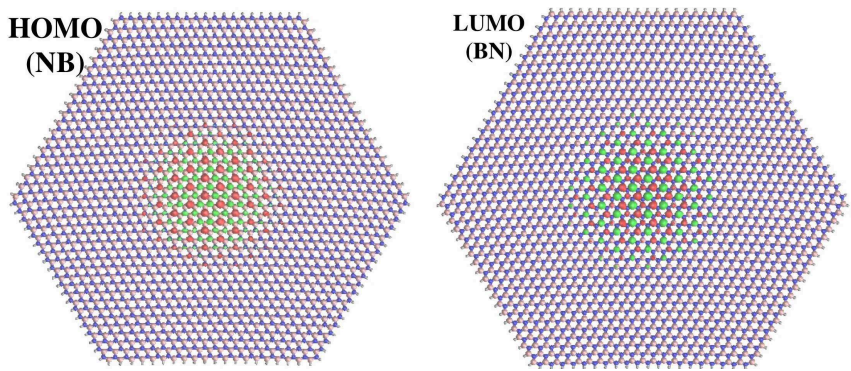}
 \caption{  \label{figLARGE}
HOMO of NB (left) and LUMO of BN (right) flakes with 2520 atoms
($d_0=4.6$~nm) subjected to the strain $\epsilon=0.15$.
\markup{Notice that for this lager system we can apply larger strain
as compared to the smaller system shown in Figure~4.} }
\end{center}
\end{figure}

In order to ensure that the observed effect is independent of the
flake size, we performed calculations~\cite{dftb+,matsci} for a larger flake with 2520
atoms ($d_0=4.6$~nm) and found similar localized
frontier orbitals as shown in \ref{figLARGE}.
 The reason for the specific spatial localization of the frontier orbitals is the rehybridization
of the electronic orbitals due to the new position of the atoms. The
induced inhomogeneous strain changes the bonds non-uniformly and
yields local dipoles which are mainly oriented radially. Note that
in general, finite flakes or nanoribbons of h-BN might be polarized
due to their finite size~\cite{PRBJavad2012}. However, here the
finite flake has zero total dipole moment because of the symmetry of
the flake even when it is subject to triaxial stain.

\subsection{Strain energy and gap variation}
 
\begin{figure}
\begin{center}
\includegraphics[width=0.45\linewidth]{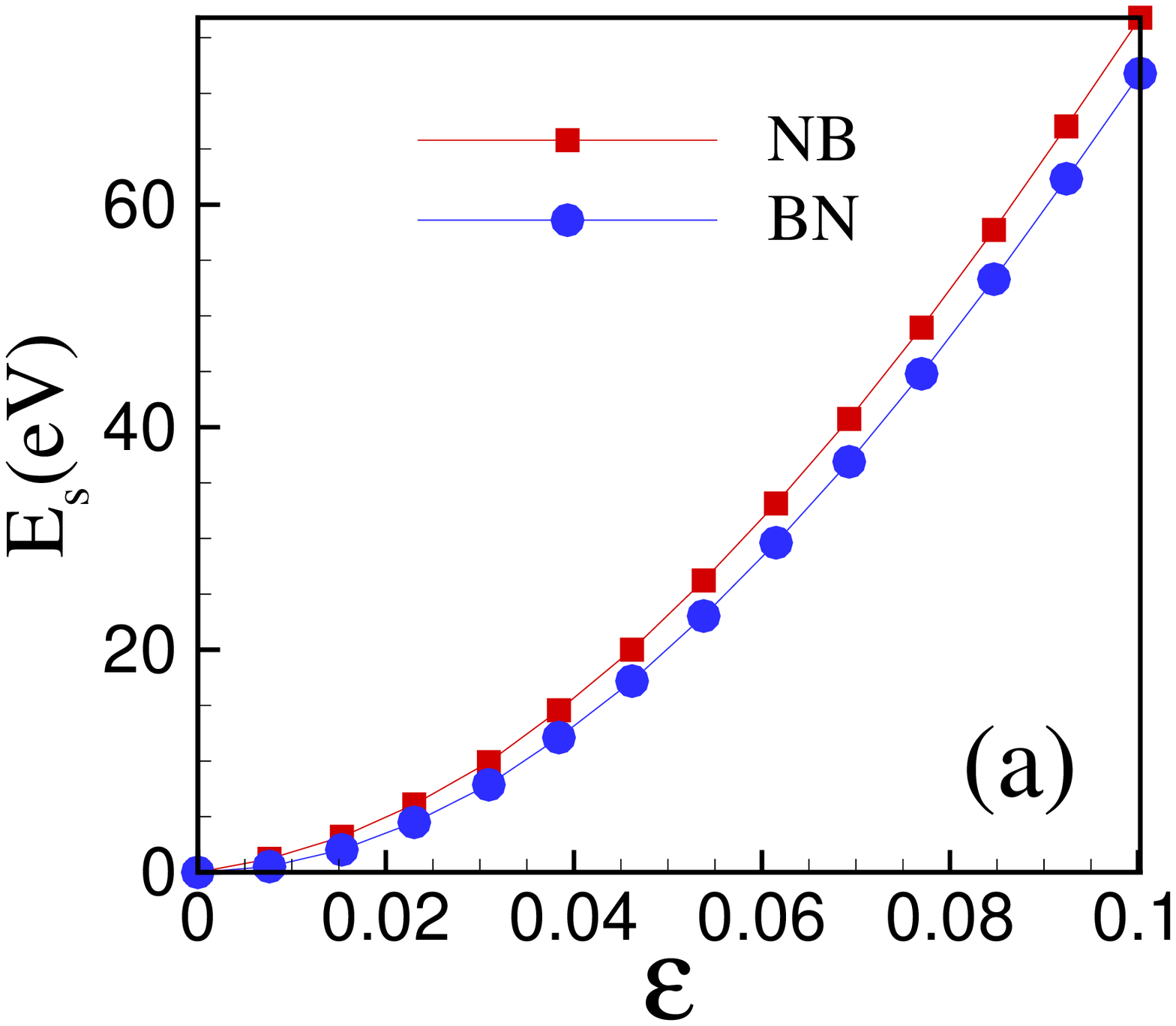}
\includegraphics[width=0.45\linewidth]{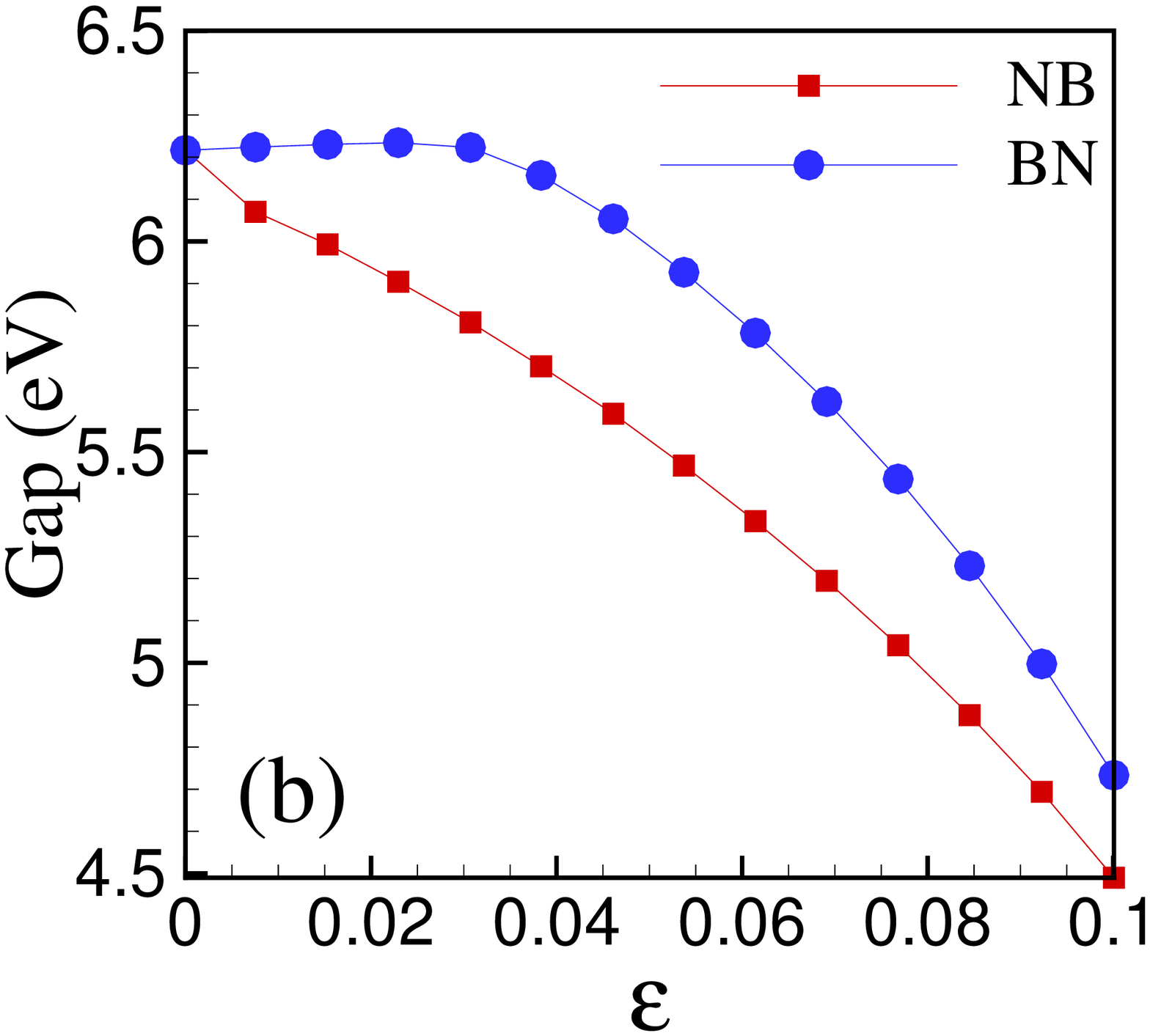}
\caption{  (a) The strain energy of the hexagonal
shaped BN and NB flakes as function of the strain parameter, i.e.
$\epsilon$. (b) The variation of the energy gap with $\epsilon$.
Each  flake consists of 252 atoms. \label{figGAP}}
\end{center}
\end{figure}

\ref{figGAP}(a) shows the variation of
the strain energy as function of the applied strain which exhibits a
quadratic behavior as expected from Hooke's law.
\ref{figGAP}(b) shows the variation of the HOMO-LUMO energy
gap with $\epsilon$.
 \markup{The density of states (DOS) spectra for strained
($\epsilon=10\%$) and unstrained  BN flakes
are shown in \ref{figNEW}(c).
By applying the strain, 
the HOMO-LUMO gap decreases by about 2~eV, and new peaks appear
above the LUMO.
%close to the HOMO  which corresponds to the
%more spread edges states in the strained BN flake, see last panel in
%\ref{figESP}(b).
}
The decreasing of the gap and modification of the DOS profile with increasing strain is
attributed to the spatial localization of the HOMO and LUMO on
regions with different ESP. For example, applying strain on the BN
flake localizes the LUMO at the center of the flake where the
electrostatic energy of an electron is lower because of high ESP.
Similarly,  applying strain on the NB flake rises its HOMO energy
level because it localizes the HOMO at the center where ESP is lower
in this case. \markup{For larger strains the
energy difference of the localized states at the center
and the edges becomes larger resulting in a smaller energy
difference of the frontier orbitals, i.e. smaller gap. The different
dependence of the gap on the strain for BN and NB (see
\ref{figGAP}(b)) is a consequence of the distributed HOMO
along all the edges and mostly the corners of the strained BN system
and the fact that the LUMO is localized at the B edges in the NB
system. } Similar effect has been seen by applying an external
electric field to nanoribbons~\cite{PRBJavad2012}. Note that the
strain induced change in the conductance of graphene was previously
investigated~\cite{naturephys2010,nanolett2009} but at present no
similar  study is available yet for h-BN.

\newpage

\subsection{Application as a  gas nanosensor}
%Finally, it is important to note that
\markup{Two alternative  methods to realize this kind of triaxial
stretching of two dimensional materials\cite{3,naturephys2010} are proposed recently.
It was reported experimentally~\cite{3} that nanobubbles of graphene
grown on a Pt(111) surface suffice very high inhomogeneous triaxial
strain which changes significantly the electronic properties of
graphene resulting in e.g. a pseudo-magnetic fields larger than 300
Tesla. The following experimental set up might also be possible:
Fixing the 2D layer (here h-BN) on a triangular shaped trench and
subsequently injecting a high pressure gas into the hole will
stretch the 2D layer and exerts a triaxial inhomogeneous strain on
the flake.}

\begin{figure*}
\begin{center}
\includegraphics[width=1\linewidth]{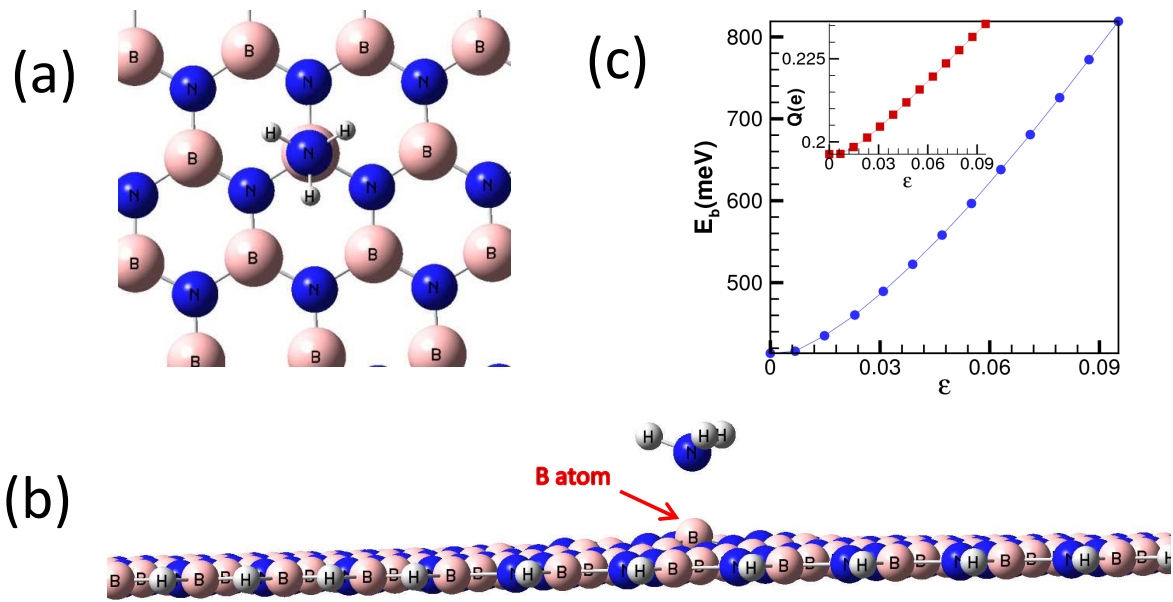}
\caption{  Top (a) and side (b) view of the adsorbed
NH$_3$ on a h-BN hexagonal flake. (c) The variation of the binding
energy and Mulliken charge on the NH$_3$ molecule (inset) vs the
strain. \label{figNH3}}
\end{center}
\end{figure*}

The controllable localizing of the frontier orbitals in the central
part of the h-BN flake is important for nanosensoring technological
applications, e.g. for filtering  gas adsorbates. The key idea is to
control  the binding energy of a molecule, namely
\begin{equation}
%E_{ad}=E_{NH_3/flake}-(E_{NH_3}+E_{flake})
E_b=(E_{molecule}+E_{flake})-E_{molecule/flake}
\end{equation}
via the applied strain.
Here  $E_{molecule}$ and $E_{flake}$ are the %ground state
energies of the pristine molecule and flake, respectively,
 while  $E_{molecule/flake}$ is the %ground state
energy of the molecule over the examined flake.
As an example, we study here %by using DFT
the adsorption of an ammonia molecule as function of the strain on
the h-BN flake. We put the NH$_3$ molecule
 in the central region of the stretched flake
and relax the system under external triaxial strain. Starting from
different orientations, we found the minimum energy when the
molecule is adsorbed onto a B atom in the middle as shown in
\ref{figNH3}(a,b). More importantly, as seen in
\ref{figNH3}(c), the binding energy depends strongly on the
strain such that by applying a strain of 10$\%$ the binding energy
is almost doubled. One notes that $\epsilon=0$ indeed gives the
minimal $E_{b}$. We  performed similar calculations when the flake
is triaxially compressed and found that the binding energy also
increases.

\begin{figure*}
\begin{center}
\includegraphics[width=.8\linewidth]{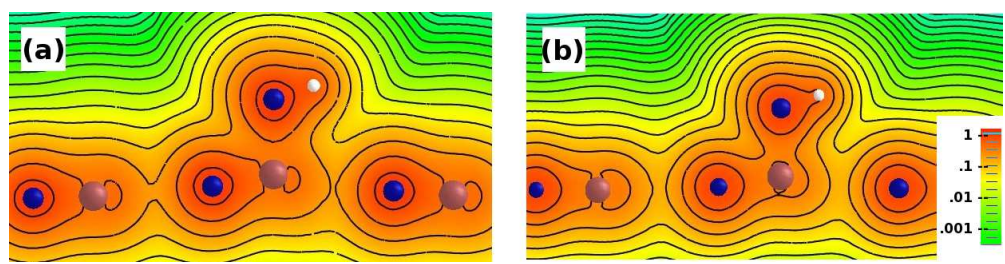}
\caption{  \label{figeden} Contour  maps of the
electron density around the chemical bond between
 an ammonia molecule and  a central B atom of a BN flake subject to zero (a) and $\epsilon=0.1$ (b) strain.
 The plane of the map
is normal to the flake and passes through N and one of the H atoms
of NH$_3$. We put N in blue, B in purple and H in white
 (c.f.  \ref{figNH3}(a)).
The successive contours differ by 10$^{1/4}$ and the color bar shows the electron density in atomic units.
}
\end{center}
\end{figure*}

NH$_3$  has one lone pair of electrons on the nitrogen atom making
it an electron donor. Therefore the molecule is positively charged
when chemically bonded to the B atom. \ref{figeden}(a) depicts
the electron density mapped on the plane containing this B-N bond when the flake is not stretched.
The impact of strain on the
redistribution of the electronic charge is seen in \ref{figeden}(b) and it predicts a
stronger covalent bonding between the molecule and the flake.
Consistently, the total Mulliken charge on NH$_3$ increases
monotonically from  0.19~e for the unstrained  flake to 0.23~e when
$\epsilon=0.1$. In the same time, the binding energy increases from
0.41~eV to 0.82~eV as shown in \ref{figNH3}(c). Triaxally
stretching the flake also quenches its buckling due to the
interaction with the adsorbed molecule such that the out of plane
height of the B atom
 under the ammonia molecule is 0.68 and 0.34~\AA~ for
zero and $\epsilon=0.1$ strains, respectively. Finally, the length
of the chemical bond between the N atom of NH$_3$ and the B atom of
the flake  also decreases monotonically form 1.83 to 1.75~\AA~
between the mentioned strains. \markup{Notice that, the binding
energy of the NH$_3$ molecule is almost the same on any B atom in
the region around the center of the BN system where the LUMO is
extended.}

The enhancement of the adsorption of NH$_3$ on the BN flake by strain %is due to the fact that
can  be explained in the framework of the so-called frontier
molecular orbitals theory~\cite{Fukui} which implies that there is a
higher tendency for adsorption of such a  molecule onto those sites
of the flake where the LUMO is localized, i.e. the central part.
However, for the NB flake  the central part has no contribution to
the LUMO and we found that the binding energy is more than twice
smaller than for the BN flake.

~~~~~
~~~~~

\section{Conclusions}
In summary, by using DFT calculations we showed that the occupied
(unoccupied) orbitals of a hexagonal shaped h-BN flake can be
localized in the center of the flake by applying triaxial strain on
the N(B) atoms at the edges of the sample.  The h-BN flake is
locally polarized but the net polarization is zero. As an example we
investigated the adsorption of ammonia and found its adsorption on
the B-edges stretched BN flake is more likely than on the N-edges
stretched flake. This is a consequence of the specific spatial
localization of the frontier orbitals. This particular kind of
localization of the frontier orbitals might have technological
applications for the design of piezoelectric and nanosensor devices.

%%%%%%%%%%%%%%%%%%%%%%%%%%%%%%%%%%%%%%%%%%%%%%%%%%%%%%%%%
\acknowledgement
%\begin{acknowledgments}
 This work was supported by the EU-Marie Curie IIF postdoc Fellowship/299855 (for
M.N.-A.), the ESF EuroGRAPHENE project CONGRAN, the Flemish Science
Foundation (FWO-Vl) and the Methusalem Funding of the Flemish
goverment. A. S. would like to thank the Universiteit Antwerpen for
its hospitality.
%\end{acknowledgments}

%\bibliography{nanosensor} 

\break
\section{Table of Contents Image}

\includegraphics[width=\linewidth]{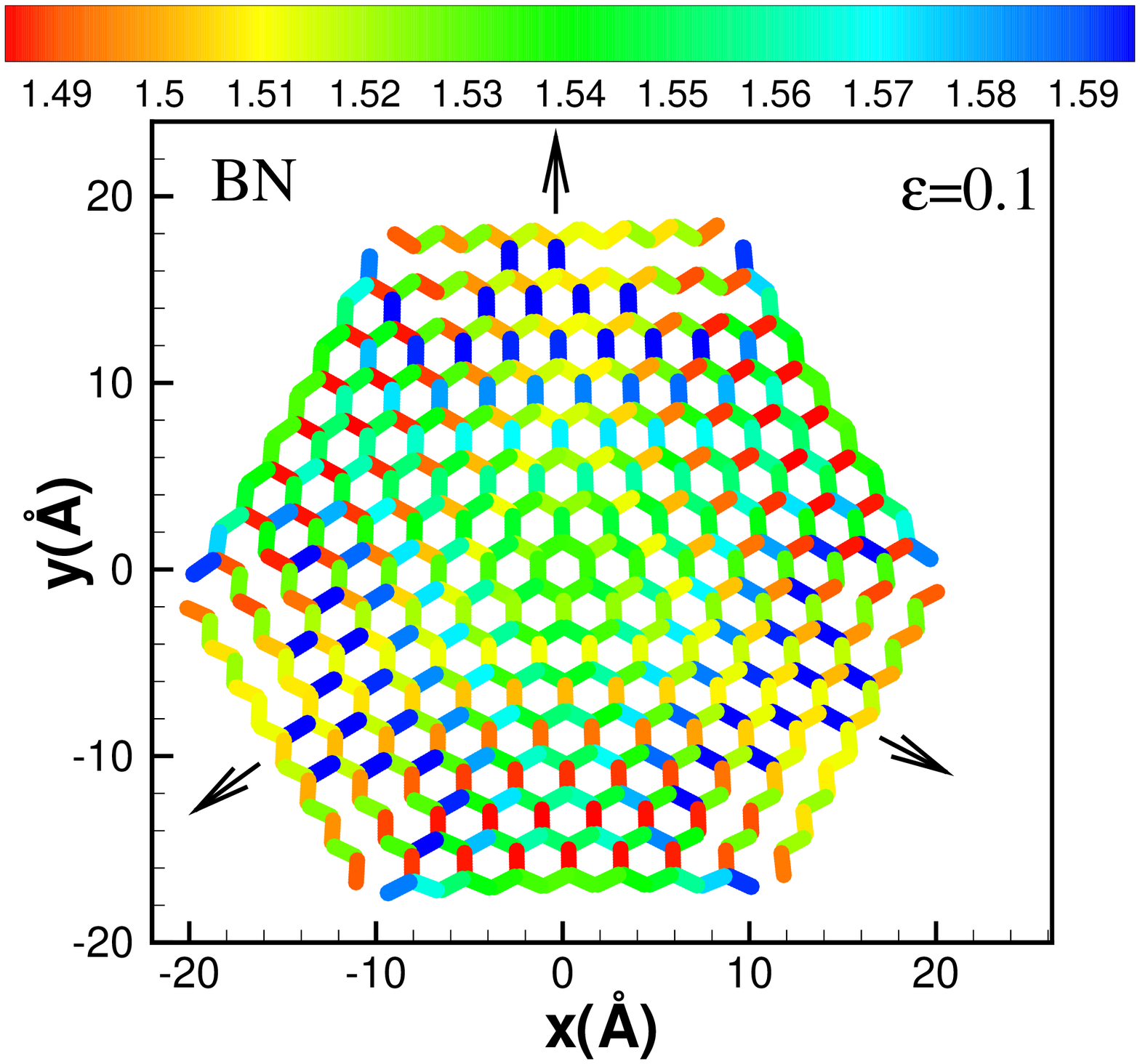}
\end{document}